\documentclass{PoS}

\title{Direct searches for the standard model Higgs boson produced in 
association with a vector boson at CDF}

\ShortTitle{Direct searches for SM Higgs boson produced in 
association with a vector boson}

\author{\speaker{Weiming Yao}\thanks{On behalf of the CDF Collaboration.}\\
        Lawrence Berkeley National Lab\\
        E-mail: \email{wmyao@lbl.gov}}

\abstract{We present the results of searches for the standard model Higgs 
boson at CDF in final states with bottom quarks. Results are derived from the
complete Tevatron Run II dataset, with a measured integrated luminosity of
9.5 fb$^{-1}$ of proton-antiproton data. The searches are performed for assumed
Higgs masses between 90 and 150 GeV, for Higgs bosons produced in
association with W or Z bosons. Employing several improved
techniques, these are currently the most sensitive searches in the world
for these processes, surpassing previous CDF results by 30\% beyond what
would be expected from the addition of new data alone. Combining the
search sensitivity of these production modes, 95\% upper confidence limits
on the standard model cross section times branching fraction are derived,
yielding an observed (expected) upper limit of 4.3 (1.8) times the
standard model prediction for a 125 GeV Higgs boson. The significance of
the data relative to the background-only hypothesis is 2.5 sigma.}

\FullConference{36th International Conference on High Energy Physics,\\
		July 4-11, 2012\\
		Melbourne, Australia}

\newcommand{\met}{ $/\!\!\!\!\!E_{T}$} 

\begin{document}

\section{Introduction}
Understanding the mechanism for electroweak symmetry 
breaking, especially by testing for the presence or absence of the standard 
model (SM) Higgs boson~\cite{higgs}, has been a major goal of particle physics 
and a central part of the Fermilab Tevatron physics program. 
Both CDF and D0 collaborations have performed direct searches for 
the standard model Higgs boson~\cite{tevcomb}. 
The new searches reported here include more data, additional 
channels, and improved analyses techniques compared to previous analyses. 
Results are derived from the complete Tevatron Run II dataset, with a measured 
integrated luminosity of 9.5 fb$^{-1}$ of proton-antiproton data. 
The searches are performed for assumed Higgs masses between 90 and 150 GeV/c$^2$.

The SM does not predict the mass of Higgs boson, $m_H$, but the global fit of the electroweak precision data, including recent 
top-quark and $W$ boson mass measurements from the Tevatron~\cite{mtop,mw},
constrains $m_H$ to be less than 152 GeV/c$^2$ at the 95\% 
conference level (CL)~\cite{ewfit}.
The direct searches from LEP~\cite{lep}, Tevatron~\cite{tevcomb}, and 
LHC results~\cite{cms,atlas} set the Higgs mass between 
116.6 and 119.4 GeV/c$^2$ or between 122.1 and 127 GeV/c$^2$ 
at the 95\% CL. Recently both LHC experiments~\cite{discoverycms, discoveryatlas} 
observed local excesses above 
the background expectations for a Higgs boson mass of approximately 125 
GeV/c$^2$. Much of the power of the LHC searches comes 
from $gg\rightarrow H$ production and Higgs boson decays to $\gamma\gamma$, 
$W^+W^-$, and $Z^+Z^-$, which probe the couplings of the Higgs boson to other 
bosons. In the allowed mass range, the Tevatron experiments are particularly 
sensitive to the associated production of the Higgs boson with a weak vector 
boson in the $b\bar b$ channel, which probes the coupling of the Higgs boson to
$b$ quarks.  

The Tevatron collider produces proton and anti-proton collision at the center
mass of 1.96 TeV with a record luminosity of 4.3~$10^{32}$~cm$^{-2}$s$^{-1}$. 
The Tevatron delivered close to 12 fb$^{-1}$ to each experiment before 
the shutdown on 30 September 2011. 
The CDF detector is a general-purpose detector, which provides excellent 
tracking, lepton identification, jets finding, missing transverse energy 
 (\met) detection, and efficient multilevel triggers. 
The details can be found elsewhere~\cite{cdf}. 

\section{Search Strategies}
The dominant Higgs production processes at the Tevatron are the 
gluon-gluon fusion ($gg\rightarrow H$) and the associated production with a $W$ or $Z$ 
boson~\cite{higgs-xsec}. The cross section for $WH$ production is twice that 
of $ZH$ and is about a factor of 10 smaller than $gg\rightarrow H$. 
The Higgs boson decay branching fraction is 
dominated by $H\rightarrow b\bar b$ for the low-mass Higgs ($m_H < 135$ 
GeV/c$^2$) and by $H\rightarrow W^+W^-$ or $Z Z^*$ for the high-mass 
Higgs ($m_H>135$ GeV/c$^2$). A search for a low-mass Higgs boson in the 
$gg\rightarrow H\rightarrow b\bar b$ channel is extremely challenging because
the $b\bar b$ QCD production rate is many orders of magnitude larger than the 
Higgs boson production rate. Requiring the leptonic decay of the associated 
$W$ or $Z$ boson greatly improves the expected signal over background ratio in 
these channels. As a result, the Higgs associated production with
$H\rightarrow b\bar b$ is the most promising channel for the low-mass Higgs
boson searches at the Tevatron.

The search strategies employed at CDF have been evolving constantly over time.  
We first maximize the signal acceptance by using efficient 
triggers, excellent lepton identifications, and powerful $b$-tagging, which 
can improve the signal to background ratio up to the 1\% level. 
Then we use multivariate analysis (MVA) to exploit the kinematic 
differences between signal and background, which can further enhance the 
signal to background ratio up to the 10\% level in the high score regions. 
The same strategies have been used to help discover the single-top and 
diboson processes at the Tevatron, and provide a solid ground for 
isolating a small signal out of a large background. 

For the $H\rightarrow b\bar b$ signatures we look for a 
$b\bar b$ mass resonance produced in association with a $W$ or $Z$ boson 
where $W$ decays into $l\nu$ or $Z$ decays into $l^+l^-$ or $\nu\bar \nu$. 
The $WH\rightarrow lvbb$ is the most sensitive channel that gives 
one high $P_T$ lepton, large \met, and two $b$-jets. Before $b$-tagging, the 
sample is predominated by the $W$ + light-flavor jets, which provides an ideal
control data to test the background modeling.

Since there are two $b$-quark jets from the low-mass Higgs decay, improving 
$b$-tagging is crucial. CDF use multivariate $b$-taggers to exploit the decay of 
long-lived $B$ hadron as displaced tracks/vertices. The typical $b$-tag efficiency is 
about 40-70\% with a mistag rate of 1-5\% per jet. 
Recently CDF combined their existing $b$-taggers into a Higgs Optimized  
$b$-tagger (HOBIT)~\cite{hobit} using a neural network tagging algorithm, 
based on sets of kinematic variables sensitive to displaced decay vertices and 
tracks within jets with large transverse impact
parameters relative to the hard-scatter vertices. 
Using an operating point which gives an equivalent rate of false tags, the new algorithm improves upon 
previous $b$-tagging efficiencies by $\approx$ 20\%.

\section{$H\rightarrow b\bar b$ Searches} 
  We describe the searches for the low-mass Higgs boson with $H\rightarrow b\bar b$ at CDF.   
\subsection{$WH\rightarrow l\nu b\bar b$} 

One of the golden channels for the low-mass Higgs boson search at the Tevatron is the Higgs 
produced in association with a $W$ boson with 
$WH\rightarrow l\nu b\bar b$~\cite{cdfwh}. We select events with one 
isolated high $P_T$ lepton (electron, muon, or isolated track), a large 
missing transverse energy, and two or three jets, of which at least one is 
required to be $b$-tagged as containing a weakly-decaying B hadron. 
Events with more than one isolated lepton are rejected. For the multivariate 
discriminant, we train a Bayesian neural network discriminant (BNN) in 
the $W$ + two or three jets for each Higgs mass, separately for each lepton type, 
jet multiplicity, $b$-tagging category. 

We perform a direct search for an excess of events in the signal
region of the final discriminant from each event category. 
Since there is no significant excess of signal events observed in the data, we set 
an upper limit at 95\% CL on the Higgs production cross section times 
branching ratio with respect to the SM predictions as a function of Higgs 
mass. CDF set an observed (expected) upper limit at 4.9(2.8) for $m_H=125$ GeV/c$^2$.

\subsection{$ZH\rightarrow l^+l^- b\bar b$}

Another interesting channel to search for the low-mass Higgs boson is 
$ZH\rightarrow l^+l^-b\bar b$~\cite{cdfllbb}. It has a clean
signature, but a low event yield due to a small branching fraction of 
$Z\rightarrow e^+e^-$ and $\mu^+\mu^-$.
We select events with two high $P_T$ leptons from $Z$ decay and
two or three jets. Events are further divided based on lepton type, 
jet multiplicity, and the number of $b$-tagged jets, similar to 
$WH\rightarrow l\nu b\bar b$. 
To increase signal acceptance we use neural networks to select loose dielectron
and dimuon candidates. The jet energies are corrected for the missing $E_T$ 
using a neural network approach. We utilize a multi-layer discriminant 
based on neural networks where separate discriminant functions are used to define
four separate regions of the final discriminant function. 
There seems to be an excess of events in the high score signal 
region, but nothing statistically significant yet. We set an observed (expected) 
upper limit at 95\% CL on the Higgs cross section times branching ratio over 
the standard model prediction at 7.2 (3.6) for the Higgs mass at 125 GeV/c$^2$. 

\subsection{$WH, ZH\rightarrow $\met$b\bar b$}

We also search for the Higgs boson in the $ZH$ and $WH$ channels where 
the $Z$ boson decays into two neutrinos or the lepton from the $W$ decay is 
undetected~\cite{cdfmetbb}. It has a large signal rate as well as a 
large QCD-multijet background. However, the final state is relatively clean, 
containing two high $E_T$ jets and a large missing transverse energy. We require
\met$> 50 $ GeV and two $b$-tagged jets. We use a track-based \met 
calculation as a discriminant against false \met.
In addition we utilize a neural network to further discriminate against the 
multi-jet background.  The final discriminant is obtained 
for a Higgs signal by combining dijet mass, track \met, and other kinematic variables.
The data are consistent with the background expectations; we set an observed (expected) upper
limit at 95\% CL on the Higgs cross section times branching ratio over the 
standard model prediction at 6.8 (3.6) for the Higgs mass at 125 GeV/c$^2$. 

\section{Combination of $H\rightarrow b\bar b$}
We combine searches for $H\rightarrow b\bar b$ in three most promising channels, and set limits with 
respect to nominal SM predictions~\cite{cdfhbb}. We combine the results using a combined likelihood 
formed from a product of likelihoods for the individual channels. Systematic uncertainties 
are treated as nuisance parameters with truncated Gaussian.
The common systematics between channels are the luminosity, 
detector related efficiencies, theoretical cross 
sections, and PDF uncertainties, which are treated as correlated.
The instrumental backgrounds are treated as independent. 
Most of nuisance parameters are well constrained by the data in the dominant background 
region and are not very sensitive to the initial-input values.
  
To validate our background modeling and search methods, we additionally perform a search for $Z\rightarrow b\bar b$
in association with a $W$ or $Z$ boson using the same final states of the SM $H\rightarrow b\bar b$ searches. 
The data sample, reconstruction, background models, uncertainties, and sub-channel divisions are
identical to those of the SM Higgs boson search, but the discriminant functions are trained using 
the signal of $Z\rightarrow b\bar b$ instead. The measured cross section of $WZ+ZZ$ is $4.1\pm 1.3 $ pb, 
which is consistent with the SM prediction of $4.4\pm 0.3$ pb.    

\begin{figure}
\begin{center}
\includegraphics[width=.48\textwidth]{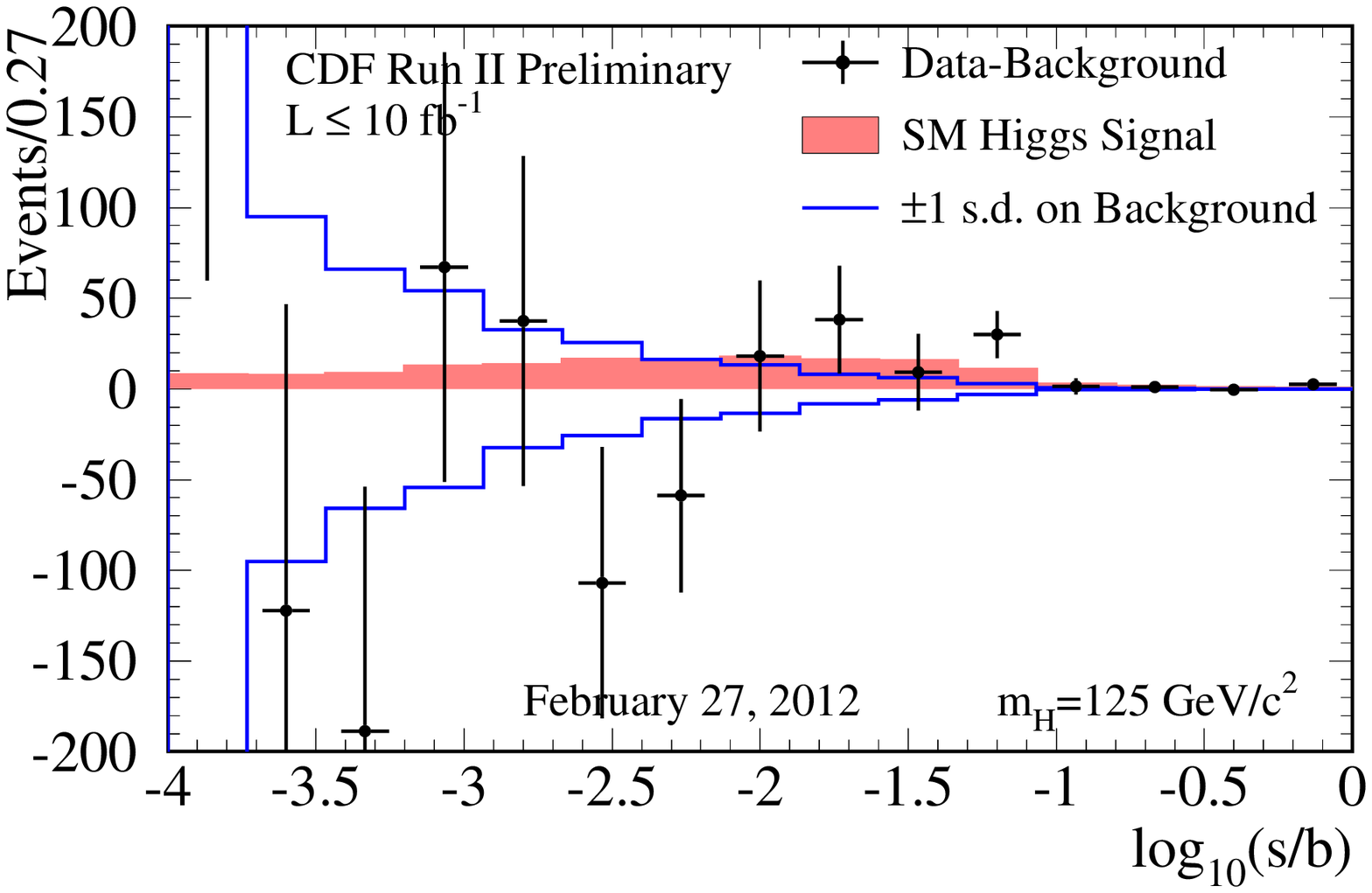}
\includegraphics[width=.48\textwidth]{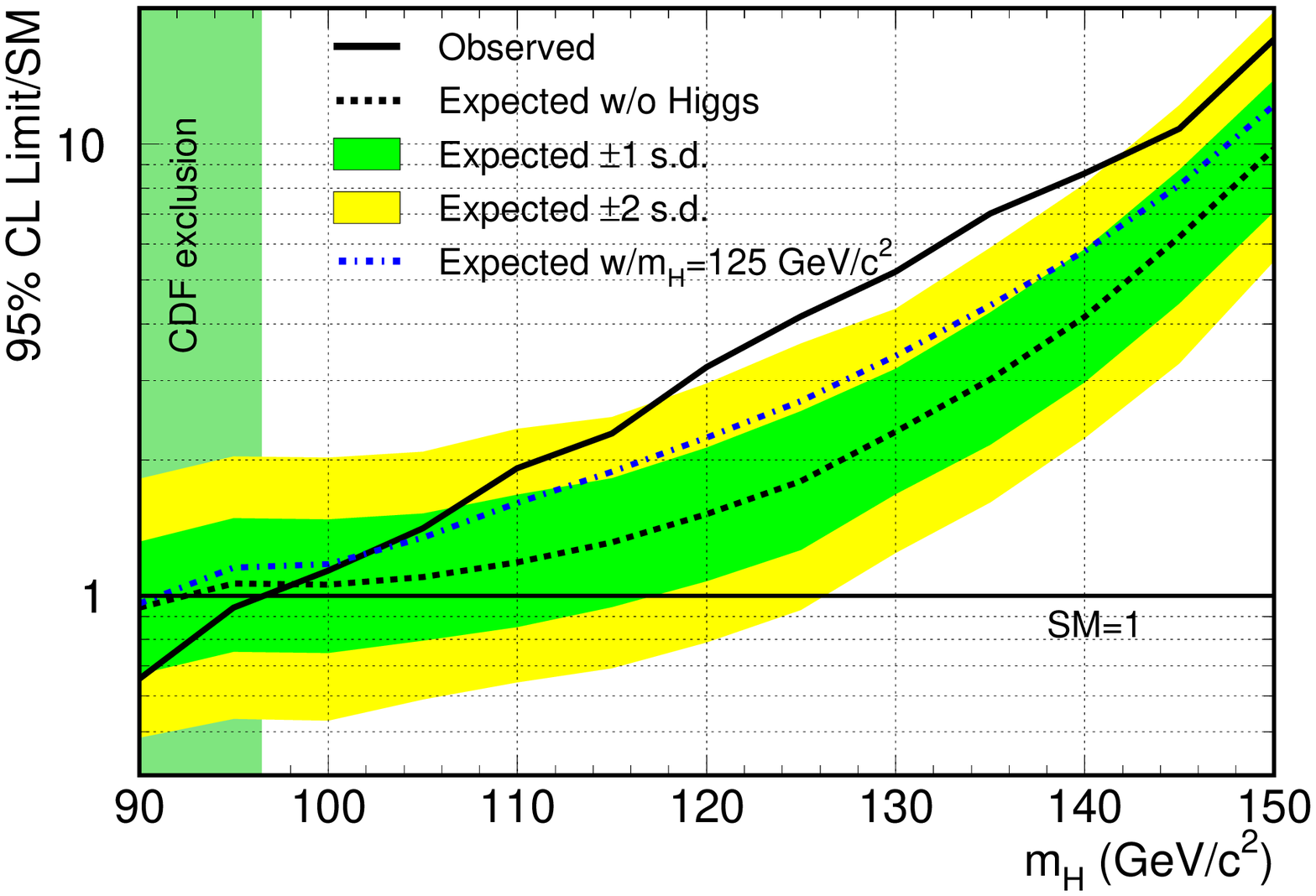}
\end{center}
\caption{(Left) Background-subtracted data distributions for the discriminant histograms, summed
for bins with similar s/b, for $m_H=125$ GeV/c$^2$. (Right) Observed and expected 95\% CL upper limits on SM Higgs 
boson production cross section with respect to SM predictions as a function of $m_H$. The shaded bands indicate the one or two sigma 
regions where the limit is expected to fluctuate, in the absence of signal. The dot-dashes are the expected limits
with the SM Higgs signal of $m_H=125$ GeV/c$^2$.} 
\label{fig:sb}
\end{figure}

To further check the consistency between data and background, 
we rebin the final discriminant from each channel in terms of the signal to background ratio (s/b) 
so that the data with similar s/b may be added without loss of sensitivity. The resulting data distributions after 
background subtraction are shown in figure~\ref{fig:sb} (left) and compared to the expected signal of $m_H =125$.  
There seems to be a small excess of Higgs boson candidate events in the highest s/b bins relative to the background-only
expectations. 

\begin{figure}
\begin{center}
\includegraphics[width=.48\textwidth]{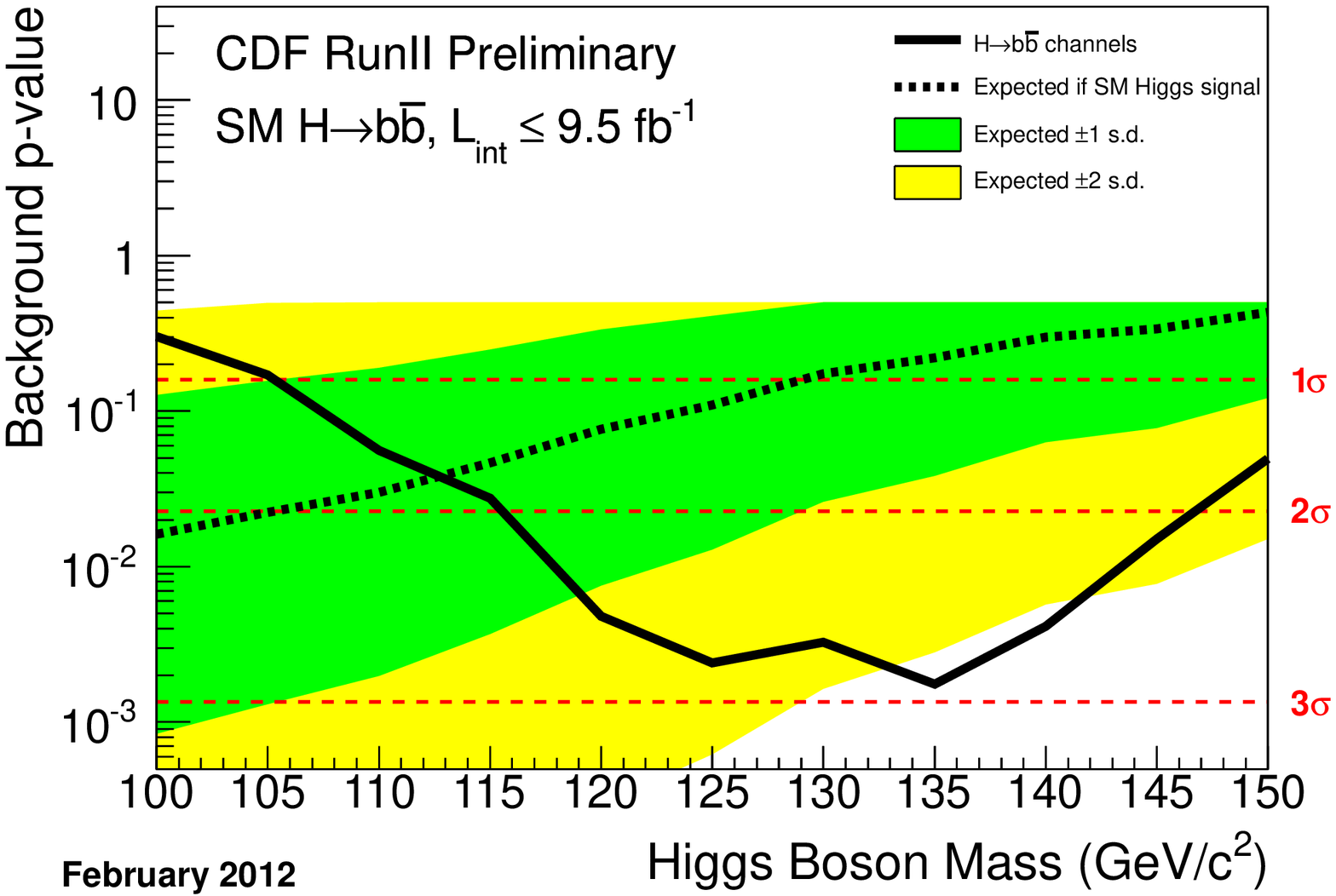}
\includegraphics[width=.48\textwidth]{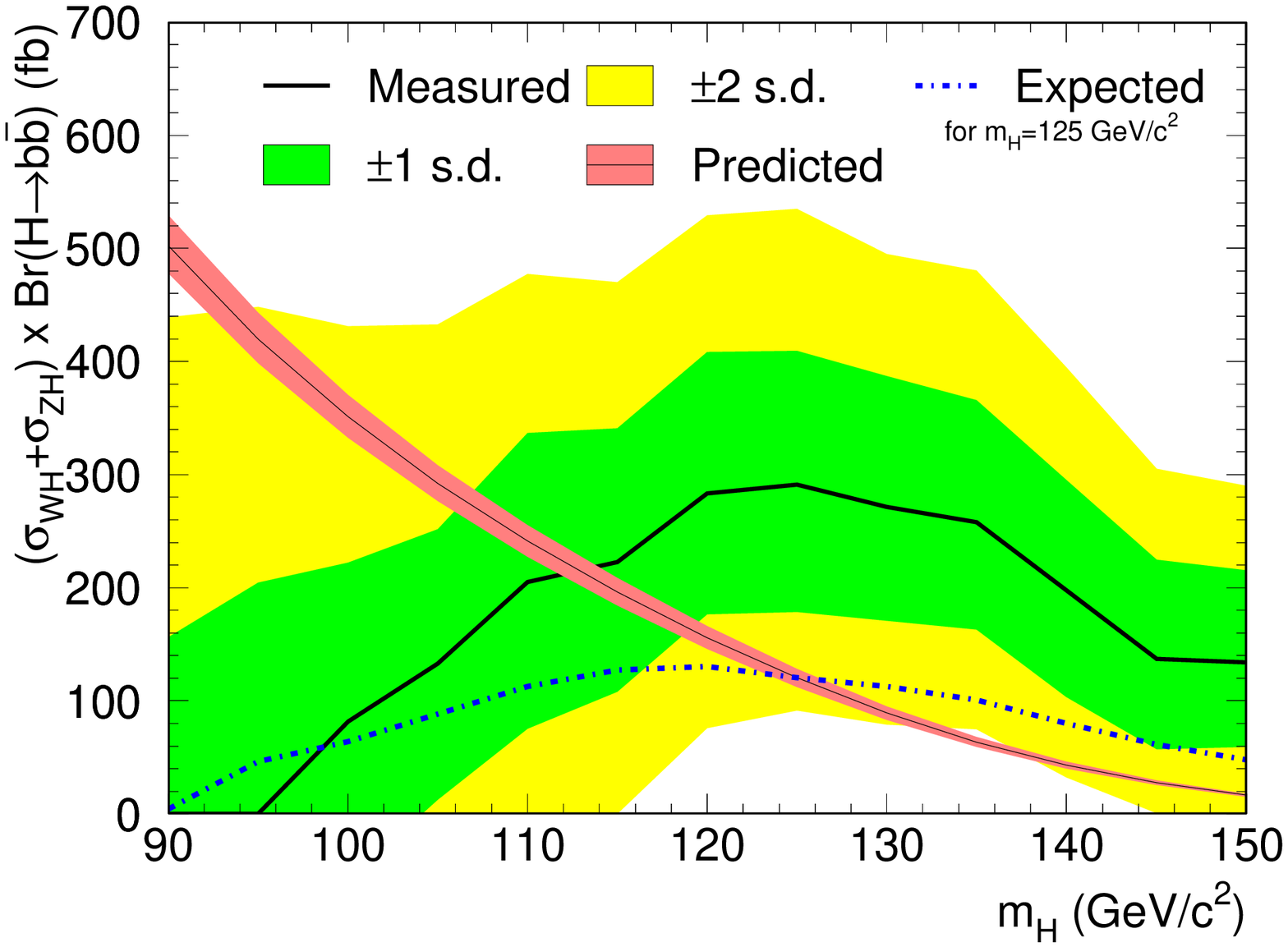}
\end{center}
\caption{(Left) Background-only p-value for the combined search and the expected value assuming a SM Higgs signal is present as 
a function of $m_H$. (Right) The best-fit cross section times branching ratio as a function of $m_H$. The shaded regions show the 
one or two sigma bands and the SM prediction is shown as a smooth falling curve. The dot-dashes are what is expected for the SM Higgs
signal of $m_H=125$ GeV/c$^2$.} 
\label{fig:pvalue}
\end{figure}

Figure~\ref{fig:sb} (right) shows the combined limit of $H\rightarrow b\bar b$ after combining three channels together.
We start to exclude the Higgs mass near the low-mass end between $90<m_H<96$ GeV/c$^2$ and 
obtain an observed (expected) limit of 4.15 (1.8) at 95\% CL on the SM Higgs production cross section with
respect to the SM prediction for $m_H=125$ GeV/c$^2$. 
There seems to be a small excess of events between 115 and 140 GeV/c$^2$.  
We quantify the excess by calculating the 
local p-value for the background-only hypothesis. Figure~\ref{fig:pvalue} (left) shows the local p-value as a function of Higgs mass 
where the solid curve is for data and the dashed curve is what is expected from 
the SM Higgs production of $m_H=125$ GeV/c$^2$. 
At $m_H=135$ GeV/c$^2$, we have a minimum local p-value=0.32\% or 2.7 sigma,
which corresponds to a global p-value=2.5 sigma including a LEE factor of 2.
Given the excess we also fit the signal cross section times the decay branching ration 
as a function of Higgs mass as shown in Figure~\ref{fig:pvalue} (right). We measure an associated production
cross section times the decay branching ratio of $(\sigma_{WH}+\sigma_{ZH})\times B(H\rightarrow b\bar b)=291^{+118}_{-113}$ (stat+sys) fb,
which is consistent with the SM prediction of $120\pm 10$ fb for $m_H=125$ GeV/c$^2$.

\section{Conclusion}
In conclusion, 
with a full dataset and many years of hard work, the CDF Collaboration has 
finally exceeded their most optimistic sensitivity projection based on 
2007 summer results. The searches are conducted for a Higgs boson that is 
produced in association with a $W$ or $Z$ boson, has a mass in the range 
90-150 GeV/c$^2$, and decays into $b\bar b$ pair. We observe an excess of 
events in the data compared with the 
background predictions, which is most significant in the mass range 
between 115 and 135 GeV/c$^2$, consistent with 
the Higgs-like particle recently observed by ATLAS and CMS. 
The most significant local 
excess is 2.7 standard deviations, corresponding to a global significance
of 2.5 standard deviations. This result may provide the first evidence for 
the Higgs coupling to $b$ quarks and we are looking forward to the discovery of 
$H\rightarrow b\bar b$ at LHC.

\end{document}